\documentclass[twocolumn,pra,nofootinbib,superscriptaddress,aps,floatfix]{revtex4-1}
\usepackage{graphicx}
\usepackage{bm}
\usepackage{amsmath}
\usepackage{color}
\def\la{\langle}
\def\ra{\rangle}
\def\capsubsz{\scriptscriptstyle}
\def\submF{_{\capsubsz F}}
\def\mF{\ensuremath{m\submF}}
\def\ket#1{{| #1 \rangle}}
\def\gJ{\ensuremath{g\submJ}}
\def\gI{\ensuremath{g\submI}}
\def\subB{_\mathrm{\capsubsz B}}
\def\muB{\ensuremath{\mu\subB}}
\def\submJ{_{\capsubsz J}}
\def\submI{_{\capsubsz I}}

\def\subeff{_\mathrm{eff}} 

\begin{document}
\title{{Structural and dynamical aspects of avoided-crossing resonances in a $3$-level $\Lambda$ system}}
\author{I. Lizuain}
\affiliation{Departamento de Qu\'\i mica-F\'\i sica, Universidad del Pa\'\i s Vasco, Apdo. 644, Bilbao, Spain}
\author{J. Echanobe}
\affiliation{Departamento de Electricidad y Electr\'onica, 
UPV-EHU, Apdo. 644, 48080 Bilbao, Spain}
\author{A. Ruschhaupt}
\affiliation{Institut f\"ur Theoretische Physik, Leibniz Universit\"at
  Hannover, Appelstra\ss e 2, 30167 Hannover, Germany} 
\author{J. G. Muga}
\affiliation{Departamento de Qu\'\i mica-F\'\i sica, UPV-EHU, Apdo. 644, Bilbao, Spain} 
\author{D. A. Steck}
\affiliation{Oregon Center for Optics and Department of Physics,
1274 University of Oregon, Eugene, Oregon 97403-1274, USA
}
\begin{abstract}
In a recent publication [Phys.\ Rev.\ A {\bf 79}, 065602 (2009)] it was shown that  an avoided-crossing resonance can be defined in different ways, according to level-structural or dynamical aspects, which do not coincide in general. 
Here a simple $3$-level system in a $\Lambda$ configuration is discussed, where the difference between both definitions of the resonance may be  observed.  
We also discuss the details of a proposed experiment to observe this difference,
using microwave fields coupling hyperfine magnetic sublevels in alkali atoms.
\end{abstract}
\maketitle

\section{Introduction}
The concept of ``resonance'' is ubiquitous in physics but  it 
is not always sharply or uniquely defined. The definition of a resonance in quantum scattering systems, for example, has been subject of endless debates among supporters of complex plane poles, of a phase shift jump, or of other criteria. 
Typically a resonance implies variations of different characteristic properties with respect to one parameter within the resonance width, but 
the extremal points 
for the different properties do not necessarily coincide;
it may even be the case that  
the variation of a property is completely missing, and the extrema 
for several of them  may be shifted with respect to each other,  
which could lead to noticeable errors if the criterion chosen for selecting the parameter value is not the most appropriate.
Resonances appear in summary as  
multifaceted phenomena, and a full characterization of their various aspects is important to control and optimize specific effects.

In a recent publication \cite{LHM08}, we have studied the definition of a resonance in quantum systems with discrete energy levels, in particular those resonances associated with avoided crossings. The crossing/avoided crossing scenario is quite common in many fields of nuclear, atomic, or molecular physics such as laser driven trapped ions \cite{CBZ94,LM07}, two level atoms coupled to a cavity mode \cite{cohen73,CDG98}, or diamagnetic hydrogen in magnetic fields \cite{WDW98}. In the avoided crossing regions, two eigenvalues of the system approach as a parameter of the system $\lambda$ is varied but then veer from each other. For  a zeroth order Hamiltonian defining the bare levels, the levels do cross at a reference value $\lambda_0$, but a perturbation connecting them causes the splitting. 
The eigenvalues also interchange their character so that, in a continuous adiabatic passage following one of the eigenvalues through the region, the system suffers a significant transformation, being dominated by different bare levels on both sides of the crossing. The resonance is also characterized by maximal oscillations for transition probabilities among the bare levels. As it was shown in \cite{LHM08}, the  parameter values of minimal splitting and of maximal transition probability do not coincide in  general, defining in this way different, structural and dynamical aspects of the resonance.

In this paper, we propose a simple physical setting, a 3-level system subjected to a 2-photon transition, where this phenomenon 
may be observed. In Sec. \ref{model_section} the model and the different 
definitions of the resonances are presented in detail.
The experimental realization is discussed in Sec. \ref{exp_proposal}, 
and the paper 
ends with a summary and a technical appendix.   

%
\begin{figure}[t!]
\begin{center}
\includegraphics[width=7cm]{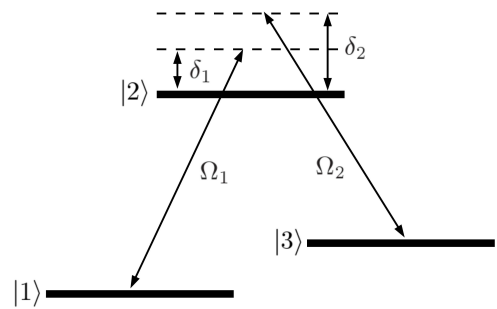}
\caption{Simple Raman 3-level setup with energy levels $1$, $2$ and $3$, detunings of the two lasers with respect to atomic transitions and coupling strengths (Rabi frequencies) $\Omega_1$ and $\Omega_2$.}
\label{raman_setup_fig}
\end{center}
\end{figure}
%

\section{The model}
\label{model_section}
\begin{figure}[t!]
\begin{center}
\includegraphics[width=7cm]{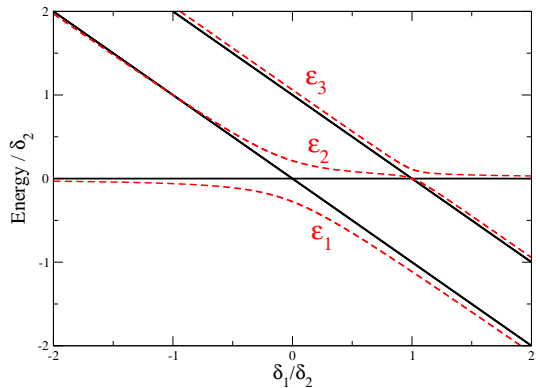}
\caption{(color online) Bare (black solid line) and dressed (red dashed line,
  $\Omega_1=\Omega_2=0.5 \delta_2$) energy levels as a function of
  $\delta_1/\delta_2$.}
\label{raman_levels_fig}
\end{center}
\end{figure}

Consider a 3-level system in a $\Lambda$-configuration (Raman 2-photon setup, Fig.~\ref{raman_setup_fig}) which is described, in a laser adapted interaction picture, by the time independent Hamiltonian ($\hbar=1$) \cite{MECL08}
\begin{eqnarray}
\label{3D_ham}
H&=& -\delta_1|2\ra\la 2|+(\delta_2-\delta_1)|3\ra\la 3|\nonumber\\
&+&\frac{\Omega_1}{2}\left(|2\ra\la1|+|1\ra\la2|\right)\nonumber\\
&+&\frac{\Omega_2}{2}\left(|3\ra\la2|+|2\ra\la3|\right)\nonumber\\
&=&\frac{1}{2}\left(\begin{array}{ccc}
0&\Omega_1&0\\
\Omega_1&-2\delta_1&\Omega_2\\
0&\Omega_2&-2(\delta_1-\delta_2)
          \end{array}\right),
\end{eqnarray}
where $\Omega_1$ and $\Omega_2$ are the coupling strengths (Rabi frequencies) of the different transitions and $\delta_1$ and $\delta_2$ the frequency detunings as shown in Fig.~\ref{raman_setup_fig}. 
When the lasers are turned off ($\Omega_1=\Omega_2=0$), the atomic states are uncoupled and the energy levels of $H$  cross  each other at $\delta_1=0$ and $\delta_1=\delta_2$. When the coupling lasers are turned on, these crossings become avoided crossings and transitions between the involved atomic energy levels at each resonance may occur, see Fig.~\ref{raman_levels_fig}.

An analytical diagonalization of the full Hamiltonian (\ref{3D_ham}) is possible but the resulting mammoth expressions are hardly illuminating. In order to have simple formulae and gain some understanding about the different aspects of the resonances, approximations will be useful. Among the two resonances  (avoided crossings) observed in the energy spectrum of the $3$-level system we shall focus on the one at $\delta_1=\delta_2$. The distance between both resonances is $\delta_2$, and since the energy splitting of each avoided crossing is proportional  to the Rabi frequencies of the coupling lasers $\Omega_1$ and $\Omega_2$, the avoided crossings will be well isolated (leading to clean transitions) as long  as $\delta_1\sim\delta_2\gg\Omega_1,\Omega_2$.
Under this condition, the state $|2\ra$ is scarcely populated and can be adiabatically eliminated to give an effective $2$-level Hamiltonian
as shown below. 
%

%
%
%
%
\subsection{Adiabatic elimination of state $|2\ra$ and effective Hamiltonian}
Substituting the general state $|\psi\ra=\sum_{n=1}^3c_n(t)|n\ra$
into the Schr\"odinger equation, the equations for the time dependent  $c_n(t)$ amplitudes are 
\begin{eqnarray}
i\dot c_1&=&\frac{\Omega_1}{2}c_2,\\
i\dot c_2&=&\frac{\Omega_1}{2}c_1-\delta_1c_2+\frac{\Omega_2}{2}c_3,\\
i\dot c_3&=&\frac{\Omega_2}{2}c_2+(\delta_2-\delta_1)c_3.
\end{eqnarray}
The usual, adiabatic-elimination argument\footnote{A more accurate and systematic theory, where the exact energy levels are obtained by iteration, may be used, see \cite{cohen73} and Appendix \ref{cohen_appendix}.} is that when $\delta_1\sim\delta_2\gg\Omega_1,\Omega_2$, the population in level $2$ remains small, nearly zero, and thus $\dot c_2(t)\approx 0$. In this way, we may write $c_2(t)$ as a function of $c_1$ and $c_3$ in the second equation and substitute in the other two, 
\begin{eqnarray}
i\dot c_1&=&\frac{\Omega_1^2}{4\delta_1}c_1+\frac{\Omega_1\Omega_2}{4\delta_1}c_3,
\\
i\dot c_3&=&\frac{\Omega_1\Omega_2}{4\delta_1}c_1+\left(\frac{\Omega_2^2}{4\delta_1}+\delta_2-\delta_1\right)c_3.
\end{eqnarray}
This system corresponds to an effective $2$-level Hamiltonian
\begin{equation}
H\subeff=\left(\begin{array}{cc}
\frac{\Omega_1^2}{4\delta_1}&\frac{\Omega_1\Omega_2}{4\delta_1}\\
\frac{\Omega_1\Omega_2}{4\delta_1}&\delta_2-\delta_1+\frac{\Omega_2^2}{4\delta_1}
             \end{array}\right)
\end{equation}
or, by shifting the zero of energy to make it symmetrical,                
\begin{equation}
H\subeff=\left(\begin{array}{cc}
-\delta\subeff&\Omega\subeff\\
\Omega\subeff&\delta\subeff
             \end{array}\right),
\end{equation}
which corresponds to an effective coupling of a laser and a two-level system with an effective coupling strength $\Omega\subeff$ and an effective detuning $\delta\subeff$ given by
\begin{eqnarray}
\Omega\subeff&=&\frac{\Omega_1\Omega_2}{4\delta_1},
\\
\delta\subeff&=&\frac{1}{2}(\delta_2-\delta_1)+\frac{\Omega_2^2-\Omega_1^2}{8\delta_1}.
\end{eqnarray}
As described in \cite{LHM08}, when both the diagonal and non-diagonal terms in a two-dimensional Hamiltonian depend on the same parameter ($\delta_1$ in this simple case), the location of the resonance is not uniquely defined and it is possible to use structural and dynamical criteria to define the resonance.
%
\subsection{Structural definition of the resonance}
From the  ``structural'' perspective of  
the energy-level diagram, the resonance may be defined as the point where the distance between the two branches of the avoided crossing is a minimum. The eigenenergies of $H\subeff$ are easily calculated,  
\begin{equation}
\epsilon_\pm=\pm\sqrt{\delta\subeff^2+\left(\frac{\Omega_1\Omega_2}{4\delta_1}\right)^2},
\end{equation}
so the minimum distance is given by the condition
\begin{equation}
\frac{\partial}{\partial\delta_1}\sqrt{\delta\subeff^2+\left(\frac{\Omega_1\Omega_2}{4\delta_1}\right)^2}=0.
\end{equation}
This corresponds to a $4^{th}$ order equation, whose solution will give us the resonance position according to the structural criterion. 
For the condition $\delta_1\sim\delta_2\gg\Omega_1,\Omega_2$, this is approximately given by (up to $4^{th}$ order terms in the frequencies)
\begin{equation}
\label{str_resonance}
(\delta_1)_S\approx\delta_2+\frac{\Omega_2^2-\Omega_1^2}{4\delta_2}-\frac{\left(\Omega_2^2-\Omega_1^2\right)^2}{16\delta_2^3}
+\frac{\Omega_1^2\Omega_2^2}{4\delta_2^3}.
\end{equation}
%
%
\begin{figure}[t]
\begin{center}
\includegraphics[width=7cm]{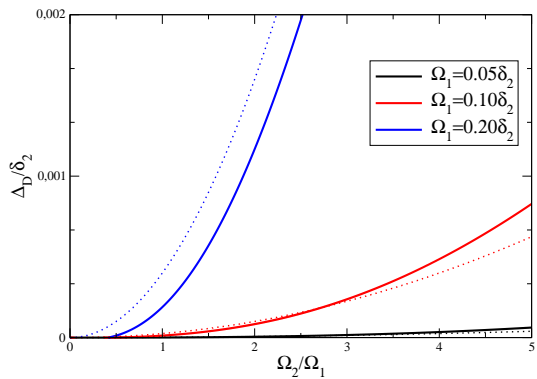}
\caption{(color online) Exact dynamical shift computed by numerically
  diagonalizing the full 3-level Hamiltonian (\ref{3D_ham}) (solid lines) as a
  function of the ratio between the coupling strengths. The approximate
  expression for the dynamical shift  (\ref{dyn_shift_approx}) is plotted with
  dotted lines.}
\label{dynamical_shift_fig}
\end{center}
\end{figure}
%

\subsection{Dynamical definition of the resonance}
{}From a dynamical perspective the resonance is defined by the value of $\delta_1$ for which the transition probability from state $|1\ra$ to state $|3\ra$ is maximum. Using $H\subeff$ this probability is easily computed,
\begin{equation}
\label{P13_dynamical}
P_{13}=\frac{\Omega\subeff^2}{\delta\subeff^2+\Omega\subeff^2}\sin^2\left(t\sqrt {\delta\subeff^2+\Omega\subeff^2}\right),
\end{equation}
and shows a maximum at $\delta\subeff=0$, which corresponds to 
\begin{eqnarray}
(\delta_1)_D&=&\frac{1}{2}\left(\delta_2+\sqrt{\delta_2^2+\Omega_2^2-\Omega_1^2}\right)\\
&\approx&\delta_2+\frac{\Omega_2^2-\Omega_1^2}{4\delta_2}-\frac{\left(\Omega_2^2-\Omega_1^2\right)^2}{16\delta_2^3}.
\label{dyn_resonance}
\end{eqnarray}
This is also the middle point where the character of each dressed energy level
changes, as discussed in \cite{LHM08}. 
The eigenstates of $H\subeff$ can be written as 
\begin{eqnarray}
|\epsilon_+\ra&=&\sin\frac{\theta}{2}|1\ra+\cos\frac{\theta}{2}|3\ra,
\\
|\epsilon_-\ra&=&\cos\frac{\theta}{2}|1\ra-\sin\frac{\theta}{2}|3\ra,
\end{eqnarray}
with $\tan\theta=-\Omega\subeff/\delta\subeff$ and real Rabi frequencies $\Omega_1$,  $\Omega_2$. The change of character of each dressed state is centered at the 
point where the linear combination has equal weights ($\theta=\pi/2$) for the states $|1\ra$ and $|3\ra$. This occurs for an effective detuning $\delta\subeff=0$;
i.e., this criterion coincides with the dynamical definition of the resonance.

Instead, the expressions (\ref{str_resonance}) and (\ref{dyn_resonance}) 
do not coincide, and are separated by a \emph{dynamical shift} $\Delta_D$ given in this approximation by 
\begin{equation}
\label{dyn_shift_approx}
\Delta_D=(\delta_1)_S- (\delta_1)_D\approx\frac{\Omega_1^2\Omega_2^2}{4\delta_2^3},
\end{equation}
which it is plotted in Fig.~\ref{dynamical_shift_fig} as a function of the
ratio between the Rabi frequencies. We see a good coincidence  between the 
exact dynamical shift and the approximation (\ref{dyn_shift_approx}) for
weak couplings. 
For strong couplings the perturbative approach breaks down, and the approximate expression deviates from the exact result.  
In any case, the simple form (\ref{dyn_shift_approx}) still gives a good estimate of the effect.

%
%
\section{Experimental determination}
\label{exp_proposal}
%
%
\begin{figure}[t]
\begin{center}
\includegraphics[width=7cm]{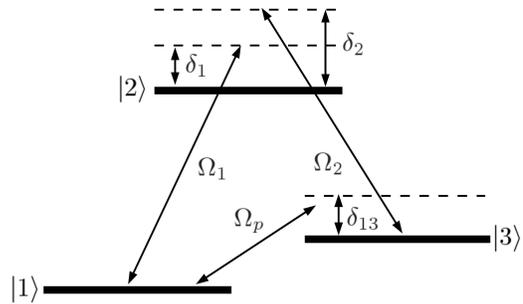}
\caption{Schematic setup with a weak probe field coupling states $|1\ra$ and $|3\ra$.}
\label{probe_fig}
\end{center}
\end{figure}
%
The ``dynamical resonance'' in Eq. (\ref{dyn_resonance}) can be easily determined experimentally by preparing the system in state $|1\ra$ for each $\delta_1$ and looking for the maximum probability of finding state $|3\ra$. The experimental determination of the minimum level splitting (``structural resonance'') requires some more work. One way is to use a third, auxiliary, weak probe field connecting states $|1\ra$ and $|3\ra$ as depicted in Fig.~\ref{probe_fig}. This transition may be an 
electric-dipole forbidden transition, such as a magnetic-dipole-allowed transition. 
Magnetic-dipole transitions are usually much weaker than electric-dipole transitions, 
a good thing in this context since we are interested in probing the dressed energy levels without 
excessively perturbing the original system.

The Hamiltonian describing the full system (including the probe field) takes the   time-dependent form 
\begin{equation}
H(t)=H+W(t),
\end{equation}
since, in general, there is no interaction picture in which the full Hamiltonian is time independent. Here, $H$ is the Hamiltonian of the original system already given in Eq. (\ref{3D_ham}), and the time-dependent perturbation is given by
\begin{eqnarray}
W(t)&=&\frac{\Omega_p}{2}\left(|3\ra\la1|e^{i\nu t}+ H.c\right),\\
\nu&=&\delta_1-\delta_2-\delta_{13},
\end{eqnarray}
$\delta_{13}$ being the detuning of the probe field with the
$|1\ra\leftrightarrow|3\ra$ transition, see Fig.~\ref{probe_fig}.

 We shall examine hereafter the resonance at $\delta_1 \approx
  \delta_2$, see Fig. \ref{raman_levels_fig}. The dressed states will be labeled
with increasing energy ($\epsilon_1<\epsilon_2<\epsilon_3$), so we have to 
measure the energy difference between $\epsilon_3$ and $\epsilon_2$ for
determining the structural resonance.

We shall 
now consider $H$ as a zeroth order Hamiltonian weakly perturbed by $W(t)$,  
and use (time-dependent) perturbation theory to obtain the transition rate from dressed state 
$|\epsilon_2\ra$
to dressed state $|\epsilon_3\ra$,  
\begin{eqnarray}
P_{|\epsilon_2\ra\rightarrow|\epsilon_3\ra}&=&\left|-i\int_0^t dt'\la\epsilon_3|W(t')|\epsilon_2\ra e^{i(\epsilon_3-\epsilon_2)t'}\right|^2\nonumber\\
&=&\Omega_p^2
\left[\frac{\alpha_{31}^2\sin^2 \frac{(\Delta\epsilon+\nu)t}{2}}{(\Delta\epsilon+\nu)^2}
+\frac{\alpha_{13}^2\sin^2 \frac{(\Delta\epsilon-\nu)t}{2}}{(\Delta\epsilon-\nu)^2}\right.\nonumber\\
&+&\left.\frac{\alpha_{13}\alpha_{31}}{\Delta\epsilon^2-\nu^2}\left(\cos^2\nu t-\cos \nu t\cos\Delta\epsilon t\right)\right],
\label{Pe2e3}
\end{eqnarray}
with $\alpha_{ij}=\la\epsilon_3|i\ra\la j|\epsilon_2\ra$ and $\Delta\epsilon(\delta_1)=\epsilon_3-\epsilon_2$. 
$P_{|\epsilon_2\ra\rightarrow|\epsilon_3\ra}$ will show peaks at $\nu\approx
\pm\Delta\epsilon$.
Thus, by changing the probe detuning $\delta_{13}$ (sweeping the value of
 $\nu$) and
   measuring the corresponding transition rate for a fixed set of parameters of the probeless system, the energy splitting
 between levels $\epsilon_2$ and $\epsilon_3$ is determined.
Following the same procedure for different values of $\delta_1$, it is possible  to find the minimum splitting and identify the structural resonance.

%
\begin{figure}[t!]
\begin{center}
\includegraphics[width=7cm]{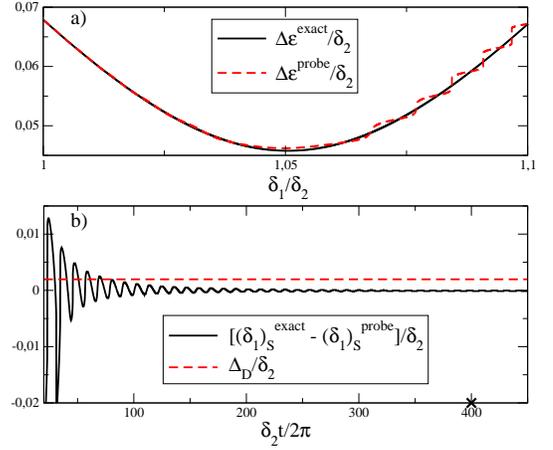}
\caption{(color online). (a) Exact energy splitting
  $\Delta\epsilon=\epsilon_3-\epsilon_2$ obtained by diagonalizing the full
  $3$-level Hamiltonian given in Eq. (\ref{3D_ham}) (black-solid line),
  compared to the energy splitting obtained by identifying one of the maxima
  (in this case, the maxima around $\nu\sim\-\Delta\epsilon$) of the
  $P_{|\epsilon_2\ra\rightarrow|\epsilon_3\ra}$ transition probability
  (red-dashed). In the $t\rightarrow\infty$ limit both lines converge (the
  calculation is done for  $\delta_2t/2\pi=125$). (b) Difference between the
  exact (probeless) position of the minimum splitting (structural resonance)
  and the structural resonance obtained from the probed system as a function
  of time (black-solid line). This difference goes to zero in the long time
  limit 
  as expected. The red-dashed line corresponds to the exact value of the
  dynamical shift $\Delta_D$ (which is indeed the precession required in the
  measurement of the structural resonance). The time marked by an $\times$ in
  (b) corresponds to $t=2\pi/\Delta_D$, the lower bound of the time
  requirement in order to resolve the dynamical shift, see
  Eq. (\ref{time_req}).  $\Omega_1=0.2\delta_2$, $\Omega_2=0.5\delta_2$.}
\label{structural_resonance_probe_fig}
\end{center}
\end{figure}
%

Note that the positions of the maxima of $P_{|\epsilon_2\ra\rightarrow|\epsilon_3\ra}$
will only be located exactly at $\nu=\pm\Delta\epsilon$ in the long time
where when $\frac{\sin((\nu\pm\Delta\epsilon) t/2)}{\nu \pm \Delta\epsilon}
\stackrel{t\to\infty}{\longrightarrow} \pi\delta(\nu\pm\Delta\epsilon)$ and the
contribution of the crossed term in Eq. (\ref{Pe2e3})
becomes negligible. At short times, the positions of the maxima are shifted due to the $\delta_1$ dependence of the $\alpha_{ij}$'s, see  Fig.~\ref{structural_resonance_probe_fig}a. To resolve the dynamical shift by this method, this effect should be smaller than the dynamical shift itself, which is indeed achieved at sufficiently long times, as shown in Fig.~\ref{structural_resonance_probe_fig}b.

As time increases the peaks of $P_{|\epsilon_2\ra\rightarrow|\epsilon_3\ra}$ become narrower (the width of each peak goes like $2\pi/t$), so in order to be able to resolve the dynamical shift, the probe beam should be applied for a time
satisfying
\begin{equation}
\label{time_req}
t\gg\frac{2\pi}{\Delta_D}.
\end{equation}
In summary, for large enough times both effects (the shift due to the $\delta_1$ dependence of the $\alpha_{ij}$'s and the width of the peaks in order to resolve the dynamical shift) can be overcome. Note also that condition (\ref{time_req}), which ensures narrow peaks, is more demanding than the times required to get rid of the shift due to the $\delta_1$ dependence of the $\alpha_{ij}$'s, see Fig.~\ref{structural_resonance_probe_fig}b.

The height of the peaks grows with time as $\sim\Omega_p^2t^2/4$, so to keep the  perturbative treatment valid, this maximum probability has to be smaller than one (weak probe field). Combining this low probe intensity condition with
the long-time condition given above,  we end up with a condition for the probe-field 
amplitude $\Omega_p$, 
\begin{equation}
\label{intensity_req}
\Omega_p\ll\frac{\Omega_1^2\Omega_2^2}{2\delta_2^3}.
\end{equation} 
%
Actually this is just an upper bound. The exact growth of the height with time is given by $(\alpha_{jk}^\pm)^2\Omega_p^2t^2/4$, but the values of the matrix elements $\alpha_{jk}^\pm$ are bounded between $0$ and $1$.
%
%
\subsection{Discussion of specific systems}
%
%
%
%
%
%
%
The most obvious setting for a Raman-transition experiment is driving optical
stimulated Raman transitions in alkali atoms.
Unfortunately, this appears to be a difficult scenario in which to study this effect.
Taking $^{87}$Rb as an example, for driving stimulated Raman
transitions between hyperfine ground levels, using lasers nearly resonant with the
D$_2$ line
($5\,^2\mathrm{S}_{1/2}\longrightarrow 5\,^2\mathrm{P}_{3/2}$ transition),
typical parameters
are a detuning $\delta_1\approx\delta_2=2\pi\cdot 10$~GHz
and Rabi frequencies $\Omega_1=\Omega_2=2\pi\cdot 200$~MHz.
These parameters give a lowest-order dynamical shift [Eq.~(\ref{dyn_shift_approx})] of
400~Hz. However, with an excited-state decay rate of $\Gamma=2\pi\cdot 6.1$~MHz,
the rate of spontaneous scattering from the Raman fields is around
$R_\mathrm{sc}\approx \Gamma(\Omega_1^{\,2}+\Omega_2^{\,2})/8\delta_1^{\,2}$,
or about 3.8~kHz.
The problem here is that the dressed states will be broadened at the kHz level,
and the interaction time of the probe will be limited, so that the
the resolution of the probe will be too poor to resolve the dynamical shift.
Decreasing the scattering rate also does not help much; for example, increasing the
detuning to $100$~GHz
leads to a scattering rate of only 38~Hz, but a dynamical shift of only 400~mHz.
The scattering rate becomes comparable to the dynamical shift for a detuning of only 1~GHz,
which is realistically too small for precision measurements.

A more promising experimental realization is possible by driving microwave
transitions in the hyperfine structure of the ground electronic level of atoms.
Here, spontaneous emission is completely ignorable, as the magnetic-dipole
transition lifetimes are much longer than any reasonable laboratory time scale.
In particular, we consider here the $n\,^{2}\mathrm{S}_{1/2}$ ground
state of alkali atoms, which is split into two hyperfine levels, 
$F=I\pm 1/2$, where $I$ is the nuclear-spin quantum number.
The three hyperfine sublevels corresponding to the setup in
Fig.~\ref{raman_setup_fig} are
$\ket{1}=\ket{F=I-1/2, \mF=-1}$ and
$\ket{3}=\ket{F=I+1/2, \mF=-1}$ for the two Raman-coupled states,
and
$\ket{2}=\ket{F=I+1/2, \mF=0}$ for the intermediate (``excited'') state.
The degeneracy of the $\ket{2}$ and $\ket{3}$ states is broken
by applying a magnetic bias field
\begin{equation}
B_\mathrm{bias} =\frac{\chi\Delta E_\mathrm{hfs}}{\muB(\gJ - \gI)} ,
\end{equation}
where $\Delta E_\mathrm{hfs}$ is the zero-field hyperfine splitting,
$\gJ$ and $\gI$ are the electronic and nuclear $g$-factors, respectively,
and $\smash{\chi:=(I+1/2)^{-1}}$.
This represents the center of an avoided crossing
of the $\ket{1}$ and $\ket{3}$ states, and thus the splitting
at this bias-field strength,
\begin{equation}
 \Delta E_{31}=
   \sqrt{1-\chi^2}\;\Delta E_\mathrm{hfs},
\end{equation}
is insensitive to first order to bias-field fluctuations.
This reduces the need for stringent experimental control over magnetic fields,
and reduces the most important systematic error in measuring the Raman resonances.
With the same magnetic field, the energy of the ``excited'' $\ket{2}$ state
is above that of the $\ket{3}$ state by an amount
\begin{equation}
 \begin{array}{l}
 \Delta E_{23}(B_\mathrm{bias}) = \gI\muB B_\mathrm{bias}
       \\ \hspace{5mm}\displaystyle {}
   +\frac{\Delta E_\mathrm{hfs}}{2} 
   \left[\sqrt{1+\chi^2} 
   -
   \sqrt{1-\chi^2} \, \right].
  \end{array}
\end{equation}
For example, for $^{87}$Rb, with 
$\Delta E_\mathrm{hfs}=h\cdot 6.835$~GHz
and $I=3/2$, the bias field is
$B_\mathrm{bias}=1.219$~kG,
and the splittings are
$\Delta E_{31} = h\cdot5.919$~GHz for the (nominal) Raman resonance, and 
$\Delta E_{23} = h\cdot 860$~MHz for the $\Omega_2$ driving transition.
The remaining ($\Omega_1$) driving transition is given by the sum
of the other two transition frequencies, or
$\Delta E_{21} = h\cdot 6.779$~GHz.
Both Raman driving transitions are driven by circularly polarized fields,
while the probe field is driven by linearly polarized field.

Continuing with the $^{87}$Rb example,
the Raman fields may be applied with Rabi frequencies of 
$\Omega_1=\Omega_2=2\pi\cdot 300$~kHz, corresponding
to field intensities of about 7.6~W/cm$^2$ on both transitions.
Microwave fields of this intensity, for example, have been realized around 6.8~GHz
in the near-field of an atom chip to manipulate a 
Bose--Einstein condensate of $^{87}$Rb \cite{Bohi09}.
Thus, a field of this strength for the $6.8$~GHz transition is feasible, and the 
field for the $860$~MHz transition should similarly pose no problem.
For a Raman detuning $\delta_1\sim\delta_2=2\pi\cdot 1$~MHz, 
the lowest-order dynamical shift from Eq.~(\ref{dyn_shift_approx})
is 2.0~kHz.

In the choice of parameters here, it is also convenient to have very different
Raman transition frequencies (6.8 and 0.9~GHz for the $\Omega_1$ and $\Omega_2$ fields, respectively),
to control the secondary ac Stark shifts
that we have not explicitly accounted for.
That is, for example, the 6.8~GHz $\Omega_1$ field driving the $\ket{1}\longrightarrow\ket{2}$
transition also couples the the $\ket{3}\longrightarrow\ket{2}$ transition
at $0.9$~GHz, albeit much farther off resonance.
As long as $\delta_2$ is held fixed, the Stark shift of $\ket{1}$ due to the $\Omega_2$ field
is inconsequential, as it simply causes a common shift of both structural and dynamic resonances.
However, the Stark shift of $\ket{2}$ due to the $\Omega_1$ field depends on $\delta_1$, and
thus can cause an additional contribution to the dynamical shift $\Delta_D$.
However, this effect is suppressed by the ratio of the detuning
$\delta_1$ from the $\ket{1}\longrightarrow\ket{2}$ transition to the detuning from the
$\ket{3}\longrightarrow\ket{2}$ transition.  This effect should thus be smaller than the 
lowest-order shift of $2.0$~kHz by a factor of about $10^{-4}$, and is therefore negligible.
Note also that it is important to have Raman detunings much smaller than the transition frequencies,
in order to suppress the effects of Bloch--Siegert shifts. By a similar argument,
the contribution of the Bloch--Siegert shifts should be of the same order as the 
secondary ac Stark shifts.

Note that uncertainties in the microwave frequencies are negligible on the scale of kHz,
so long as the fields are derived from digital synthesizers.
However, the splitting at each detuning must be determined to an accuracy
finer than the  $2.0$~kHz shift.
Thus, to resolve this shift of $2.0$~kHz,
the (6.8~GHz) probe beam should be applied for a 
time much longer than $500$~$\mu$s [Eq.~(\ref{time_req})], 
with a Rabi frequency small compared to $2\pi\cdot 1$ kHz [Eq.~(\ref{intensity_req})]. 
The probe field then requires a correspondingly much lower intensity, as compared to the
Raman fields.
The atoms will also need to be well-confined on ms time scales, without inducing
spontaneous emission.  Loading laser-cooled atoms into a dipole trap---formed by 
the focused light of a CO$_2$ laser---accomplishes
this confinement,
with negligible perturbation to the hyperfine structure of the ground electronic state.

Care must also be taken in preparing the atoms for the probe measurement.
Since the goal is to measure the splitting of the \textit{dressed} states
at a particular detuning, as described above, we must prepare the atoms in
only one of the dressed states.  This is effected, for example, by first
optically pumping the atoms (in the absence of the Raman fields, and with only a small
magnetic bias field of the order of $100$~mG to prevent mixing of states) into the 
$\ket{1}=\ket{F=1,\mF=-1}$ bare state.
This is accomplished by driving the 
$5\,^2\mathrm{S}_{1/2}, F=1\longrightarrow 5\,^2\mathrm{P}_{3/2}, F'=1$ optical transition
with circularly polarized light,
while optically depumping the $F=2$ ground hyperfine level.
The $1.2$~kG field should then be turned on adiabatically to produce the correct level
configuration without inducing any transitions.
The Raman fields should then also be turned on adiabatically, but far from Raman resonance.
They can then be adiabatically chirped to the desired Raman detuning,
transferring the atoms from $\ket{1}$ to $\ket{\epsilon_2}$.
The probe field should then be activated to attempt to drive atoms to the
other dressed state $\ket{\epsilon_3}$.
Finally, the Raman fields should again be detuned and adiabatically turned off,
and the magnetic field turned off adiabatically as well.
The population transferred to $\ket{\epsilon_3}$, and thus to
$\ket{3}= \ket{F=2,\mF=-1}$, is then measured by fluorescence
detection of the $F=2$ population.
Finer resolution of the Raman splitting is also possible by employing a Ramsey-interference
technique, applying the probe in two pulses separated in time.

\section{Summary}
As discussed in \cite{LHM08}, avoided crossing resonances between dressed energy levels are not uniquely defined. In this paper a $3$-level atom in a $\Lambda$ configuration has been proposed as a simple physical system where the distinction between structural and dynamical aspects of a resonance could be observed. By adiabatically eliminating the third level, an effective $2$-dimensional Hamiltonian has been obtained, from which an approximate analytical expression for the dynamical shift (between structural and dynamical resonances) has been given. 

While the dynamical resonance is in principle easy to observe by measuring the
maximum rate of a given atomic transition, the determination of the structural
resonance is more delicate. We have proposed a method consisting of a weak
probe field.
In order to resolve the dynamical shift by this method, a low-intensity probe field 
must be applied for sufficiently long times. An ideal setting is 
provided by  microwave transitions in the hyperfine structure of the ground state level of alkali atoms. 
%
%
\begin{acknowledgments}
We are very grateful to C. Cohen-Tannoudji for useful comments. We acknowledge    
support by Ministerio de Innovaci\'on y Ciencia (FIS2009-12773-C02-01), 
Basque Government Grant IT472-10, and
National Science Foundation Grant PHY-0855412.
\end{acknowledgments}
\appendix
\section{Resolvent method}
\label{cohen_appendix}

It is possible to improve systematically the adiabatic approximation by using a more accurate method, where the energy levels are given exactly by an implicit Hamiltonian, \cite{cohen73,CDG98}. In order to use this resolvent method, we may divide the starting Hamiltonian (\ref{3D_ham}) as $H=H_0+V$ with
\begin{eqnarray}
 H_0&=&\left(\begin{array}{ccc}
0&0&0\\
0&-\delta_1&0\\
0&0&\delta_2-\delta_1
          \end{array}\right)\\
V&=&\frac{1}{2}\left(\begin{array}{ccc}
0&\Omega_1&0\\
\Omega_1&0&\Omega_2\\
0&\Omega_2&0
          \end{array}\right).
\end{eqnarray}
The bare energy levels (eigenergies of $H_0$) corresponding to the bare states $|1\ra$, $|2\ra$ and $|3\ra$ are given respectively by
\begin{eqnarray}
\epsilon_1^{(0)}&=&0,\\
\epsilon_2^{(0)}&=&-\delta_1,\\
\epsilon_3^{(0)}&=&\delta_2-\delta_1.
\end{eqnarray}
Around $\delta_1\approx\delta_2$, energy levels $\epsilon_1^{(0)}$ and $\epsilon_3^{(0)}$ are degenerate but will form an avoided crossing when the lasers are turned on. Around this value of $\delta_1$, the system will be described by an implicit effective $2$D Hamiltonian \cite{cohen73},
\begin{eqnarray}
\label{Heff_not_symmetric}
H\subeff&=&\left(\begin{array}{cc}
\epsilon_1^{(0)}+R_{11}&R_{13}\\
R_{31}&\epsilon_3^{(0)}+R_{33}
              \end{array}\right),
\end{eqnarray}
where $R_{ij}=\la i|R|j\ra$ are the matrix elements of the level shift operator $R$,
\begin{eqnarray}
R(E)&=&\sum_{n=0}^{\infty}PV\left(\frac{Q}{E-H_0}V\right)^nP,
\end{eqnarray}
with $P=|1\ra\la1|+|3\ra\la3|$ and $Q=1-P=|2\ra\la2|$. The effective Hamiltonian (\ref{Heff_not_symmetric}) can be written in a symmetrical way by changing the zero of the energy
\begin{eqnarray}
H\subeff&=&\left(\begin{array}{cc}
-\delta\subeff&R_{13}\\
R_{31}&\delta\subeff
               \end{array}\right)+
C\left(\begin{array}{cc}
       1&0\\
	0&1
      \end{array}\right),
\end{eqnarray}
with
\begin{eqnarray}
\delta\subeff&=&\frac{1}{2}\left(\epsilon_3^{(0)}-\epsilon_1^{(0)}+R_{33}-R_{11}\right)\\
C&=&\frac{1}{2}\left(\epsilon_1^{(0)}+\epsilon_3^{(0)}+R_{11}+R_{33}\right).
\end{eqnarray}
Explicit expressions of the needed elements are  \emph{exactly}  given by
\begin{eqnarray}
R_{11}&=&\frac{\Omega_1^2}{4(E+\delta_1)},\\
R_{33}&=&\frac{\Omega_2^2}{4(E+\delta_1)},\\
R_{13}&=&\frac{\Omega_1\Omega_2}{4(E+\delta_1)}=R_{31},
\end{eqnarray}
where we have taken into account that $\epsilon_2^{(0)}=-\delta_1$, 
and  
\begin{eqnarray} \delta\subeff&=&\frac{1}{2}\left[\delta_2-\delta_1+\frac{\Omega_2^2-\Omega_1^2}{4(E+\delta_1)}\right],
\\
C&=&\frac{1}{2}\left[\delta_2-\delta_1+\frac{\Omega_2^2+\Omega_1^2}{4(E+\delta_1)}\right].
\end{eqnarray}
Note that the elements of the implicit Hamiltonian depend on the eigenenergy $E$. 
Thus the energy levels will be obtained by iteration. Once the energy levels are calculated (with a given precision) the different resonance loci will be calculated as described above. Choosing $E=0$ for the first iteration corresponds exactly to the adiabatic elimination approximation in Sec. \ref{model_section}.


\end{document}